\documentclass[12pt,a4paper,final]{iopart}

\usepackage{cite}
\usepackage{iopams}
\usepackage{graphicx}
\usepackage[breaklinks=true,colorlinks=true,linkcolor=blue,urlcolor=blue,citecolor=blue]{hyperref}

\begin{document}

\title[]{\bf Field representation of a watt balance magnet by partial profile measurements}

\author{S. Li$^{1,2*}$, J. Yuan$^{1}$, W. Zhao$^{1}$, S. Huang$^{1}$}
\address{$^1$Department of Electrical Engineering, Tsinghua University, Beijing 100084, China}
\address{$^2$National Institute of Metrology, Beijing 100029, China}
\ead{leeshisong@sina.com}

\begin{abstract}
Permanent magnets with high-permeability yokes have been widely used in watt balances for supplying a robust and strong magnetic field at the coil position. Subjected to the mechanical realization, only several $B_r(z)$ (radial magnetic field along the vertical direction) profiles can be measured by coils for field characterization. In this article, we present an algorithm that can construct the global magnetic field of the air gap based on $N(N\geq1)$ additional measurements of $B_r(z)$ profiles. The proposed algorithm is realized by polynomially estimating the $B_z(r)$ function with analysis of basic relations between two magnetic components in air gap, i.e., $B_r(r,z)$ and $B_z(r,z)$, following the Maxwell's equations. The algorithm, verified by FEM simulations, can characterize the three-dimensional contribution of the magnetic field for a watt balance magnet with acceptable accuracy, which would supply basic field parameters for alignment and misalignment corrections.
\end{abstract}
\submitto{Metrologia}
\clearpage
\section{Introduction}
The watt balance, which was originally proposed by Dr. B P Kibble at the National Physical Laboratory (NPL, UK) in 1975 \cite{Kibble1976}, is an experiment currently for precision measurement of the Planck constant $h$, and in future for maintaining one of the seven base units, the kilogram \cite{mills2006redefinition}. The operation of a watt balance is divided into two independent measurement modes, respectively known as the weighing mode and the velocity mode. In the weighing mode, a magnetic force generated by current-carrying coil in a magnetic field is balanced by the weight of a test mass $m$ as $mg=BLI$, where $g$ denotes the local gravitational acceleration, $B$ the magnetic flux density at coil position, $L$ the wire length of the coil, and $I$ the current in coil. In the velocity mode, the coil is moved in the magnetic field with a velocity $v$ in vertical direction, inducing a voltage $U=BLv$ on the coil terminals. By a combination of the weighing and velocity modes, the geometrical factor $BL$ is eliminated and a virtual watt balance equation, i.e., $UI=mgv$, would be obtained. As a result, the Planck constant $h$ can be related to the test mass $m$ by comparing the electrical power to the mechanical power in conjunction with the quantum Hall effect \cite{klitzing1980new} and the Josephson effect \cite{josephson1962possible}. Since the proposal in 1975, the watt balance has been widely spread and persuaded at many national metrology institutes (NMIs) \cite{NPL, NIST, METAS, BIPM, LNE, NIM, MSL, NRC, KRISS}. The detailed principle and recent progress of watt balance experiments at NMIs can be found in several review papers, e.g., \cite{stockR, steiner2013history, LSS2014R}.

In order to obtain enough magnetic force, e.g., 5N, and in the meanwhile keep the power assumption of the coil as low as possible, a strong magnetic field is preferred in watt balance. In realization, the watt balance usually employs permanent magnets as the magnetic flux source and uses soft yokes to guide the generated flux through a small air gap where the coil is placed. As most of the magneto motive force (MMF) in such magnetic circuits would drop across the air gap, the magnetic flux in the air gap would be strong. Besides, the field boundary on yoke-air surfaces is sharp with high-permeability yokes, and hence the field uniformity in the air gap is good. But one disadvantage for such magnetic circuit is that the magnetic field gradient in the horizontal direction is obviously increased, which may bring difficulties for the alignment \cite{Robinson2012alignment}. As is noticed, the alignment is one of the most difficult procedures in operating a watt balance, and several techniques that can relax the alignment have been proposed: the simultaneous measurement of two operating modes has been carried out in watt balance at the Bureau International des Poids et Mesures (BIPM) \cite{BIPM}, which has been further developed by Dr. I A Robinson in \cite{Robinsontwophase}; new experimental designs of watt balances, of course, can relax the alignment \cite{KibbleWB}.

Here in this paper, a misalignment correction idea based on measurement of coil position and representation of magnetic field in the air gap, is considered. The merit of the correction method is that it would relax the alignment and can be applied on all the existing watt balances without redeveloping new apparatus. The coil position in both modes of a watt balance can be well measured as a necessary requirement of the experiment. In order to eliminate the misalignment error to a level below $1\times10^{-8}$ by correction, a key procedure is to know the three-dimensional (3D) magnetic profile in the operating interval of the air gap. However, limited to the mechanical realization, only several vertical profiles of $B_r(z)$ at different horizontal coordinates can be determined by the gradient coil (GC) method \cite{Seifert2014} or a high resolution magnetic probe. Knowing a set of subjected information, a natural question is: how can we solve the whole filed in the air gap by using minimum measurable quantities? To answer this question, in this paper, we present an algorithm that can represent the 3D magnetic profile of the air gap based on $N(N\geq1)$ additional measurements of $B_r(z)$ profiles.

The outline for the rest of this paper is organized as follows: the algorithm principle is presented in section 2 which is originated from polynomial estimation of the $B_z(r)$ function following Maxwell's equations in the static magnetic circuit, and in section 3, numeral verifications by a cubic estimation of the $B_z(r)$ function with one additional $B_r(z)$ measurement are shown; in section 4, the equations for misalignment correction are reviewed and discussed; a conclusion is drawn in section 5.

\section{Principle}
\label{section2}
\subsection{Magnet structure}
\begin{figure}
\center
\includegraphics[width=3.5in]{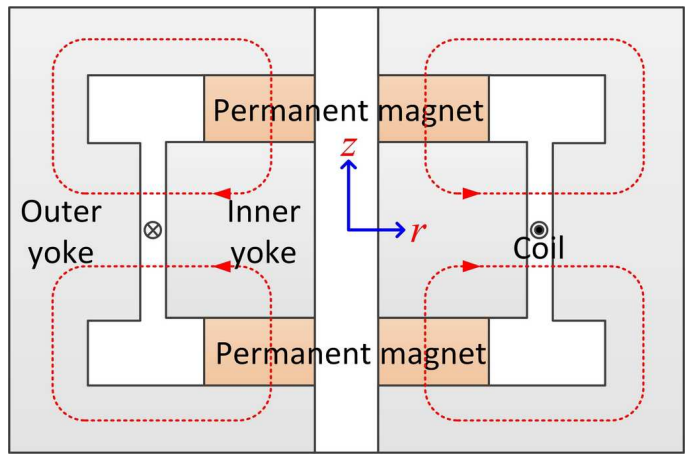}
\caption{A typical magnetic structure used in watt balances (sectional view). Two circles in the upper and lower parts of the magnet denote the main flux in the magnetic circuit. The flux direction shown is one of two opposite cases when the northern poles of permanent magnets are facing the inner yoke.}
\label{Fig.1}
\end{figure}

A typical magnetic structure employed in watt balance with one coil and two permanent magnet rings has been shown in figure \ref{Fig.1}. The magnetic circuit was first presented in the BIPM watt balance group \cite{Stock2006}, and later followed by the METAS Mark II watt balance at the Federal Institute of Metrology, Switzerland \cite{Baumann2013}, the NIST-4 watt balance at the National Institute of Standards and Technology, USA \cite{Seifert2014}, the MSL watt balance at the Measurement Standards Laboratory, New Zealand \cite{Sutton2014}, and the KRISS watt balance at the Korea Research Institute of Standards and Science, South Korea \cite{KRISS}. The shown magnet structure is preferred due to the advantages including generation of a strong magnetic field in the air gap, a good magnetic shielding, and the possibility of a compact size realization.

In the shown symmetrical magnetic structure, two permanent magnet rings are arranged with opposite magnetization poles and their flux, guided by high-permeability yokes, runs horizontally through the air gap where the coil is suspended. The air gap is located between two parallel cylindrical yokes, i.e., the inner yoke and the outer yoke, and its width is conventionally several centimeters. The height of the air gap is typically tens of centimeters and the central part with several centimeters is the applied measurement interval in the velocity mode. Normally, both the weighing and velocity modes are operated in a chosen measurement interval along the central $B_r(z)$ profile, which can be well adjusted with a relative filed change of several parts in $10^{4}$.

Note that there are some other realizations of watt balance magnets, e.g., \cite{NRC,Gournay2005}. Although the model and analysis presented in this paper is based on the magnet shown in figure \ref{Fig.1}, they can be also applied in other magnet structures.

\subsection{Magnetostatic equations}
We first present the generalized magnetostatic equations to describe the field in the air gap for further discussion of the proposed algorithm. For a static magnet circuit shown in figure \ref{Fig.1}, the differential forms of Maxwell's equations for the magnetic field in the air gap are written respectively as
\begin{equation}
\nabla\cdot {\bf B}=0,
\label{eq.Max1}
\end{equation}
\begin{equation}
\nabla\times {\bf B}=\mu_0 {\bf J},
\label{eq.Max2}
\end{equation}
where ${\bf J}$ denotes a three dimensional vector of the current density, $\bf B$ is the vector of the magnetic flux density in the air gap, and $\mu_0$ is the permeability in vacuum; $\nabla$ is the Del operator defined as $\nabla={\bf r_0}(\partial/\partial r)+{\bf \phi_0}(1/r)(\partial/\partial \phi)+{\bf z_0}(\partial/\partial z)$ where (${\bf r_0}$, ${\bf \phi_0}$, ${\bf z_0}$) is a unit vector in the cylindrical coordinate ($r$, $\phi$, $z$).

Writing the vector $\bf B$ in equation (\ref{eq.Max1}) in forms of three components in the cylindrical coordinate, i.e., ${\bf B}=(B_r,B_\phi,B_z)$, we obtain
\begin{equation}
\frac{1}{r}\frac{\partial [rB_r(r,\phi, z)]}{\partial r}+\frac{1}{r}\frac{\partial B_\phi(r,\phi, z)}{\partial \phi}+\frac{\partial B_z(r,\phi, z)}{\partial z}=0.
\label{eq.maxI}
\end{equation}
Since the presented magnet in figure \ref{Fig.1} has a $r-z$ symmetrical structure, it meets the following relations for the magnetic flux density in the air gap that
 \begin{equation}
 B_\phi=0, ~~\frac{\partial B_r(r,\phi, z)}{\partial\phi}=0, ~~\frac{\partial B_z(r,\phi, z)}{\partial\phi}=0.
 \label{eq:3zeros}
 \end{equation}
 As a result, equation (\ref{eq.maxI}) can be rewritten in a two dimensional form as
\begin{equation}
\frac{1}{r}\frac{\partial [rB_r(r,z)]}{\partial r}+\frac{\partial B_z(r,z)}{\partial z}=0.
\label{eq.maxI2D}
\end{equation}

Similarly, if ${\bf J}=(J_r,J_\phi,J_z)$, equation (\ref{eq.Max2}) can be expressed by three components in the cylindrical coordinate respectively as
\begin{equation}
\frac{1}{r}\frac{\partial B_z(r,\phi,z)}{\partial \phi}-\frac{\partial B_\phi(r,\phi,z)}{\partial z}=\mu_0J_{r}(r,\phi,z),
\label{eq.MaxII1}
\end{equation}

\begin{equation}
\frac{\partial B_r(r,\phi,z)}{\partial z}-\frac{\partial B_z(r,\phi,z)}{\partial r}=\mu_0J_{\phi}(r,\phi,z),
\label{eq.MaxII2}
\end{equation}

\begin{equation}
\frac{1}{r}\frac{\partial [rB_\phi(r,\phi,z)]}{\partial r}-\frac{1}{r}\frac{\partial B_r(r,\phi,z)}{\partial \phi}=\mu_0J_{z}(r,\phi,z).
\label{eq.MaxII3}
\end{equation}

Note that equations (\ref{eq.MaxII1})-(\ref{eq.MaxII3}) are written in the space of the air gap, where the current density should equal zero, i.e., $J_r=J_\phi=J_z=0$.
Besides, based on equation (\ref{eq:3zeros}), both left sides of equation (\ref{eq.MaxII1}) and equation (\ref{eq.MaxII3}) equal zero, therefore, equations (\ref{eq.MaxII1})-(\ref{eq.MaxII3}) can be simplified by an equation in two dimensional form as
\begin{equation}
\frac{\partial B_r(r,z)}{\partial z}-\frac{\partial B_z(r,z)}{\partial r}=0.
\label{eq.MaxII2D}
\end{equation}

Equations (\ref{eq.maxI2D}) and (\ref{eq.MaxII2D}), which present the basic relations between two components of the magnetic flux density, i.e., $B_r(r, z)$ and $B_z(r, z)$, are the magnetostatic equations for the magnetic field in the air gap of the shown watt balance magnet.

\subsection{Fringe field effect of the air gap}
\label{2.3}
\begin{figure}
\center
\includegraphics[width=3.2in]{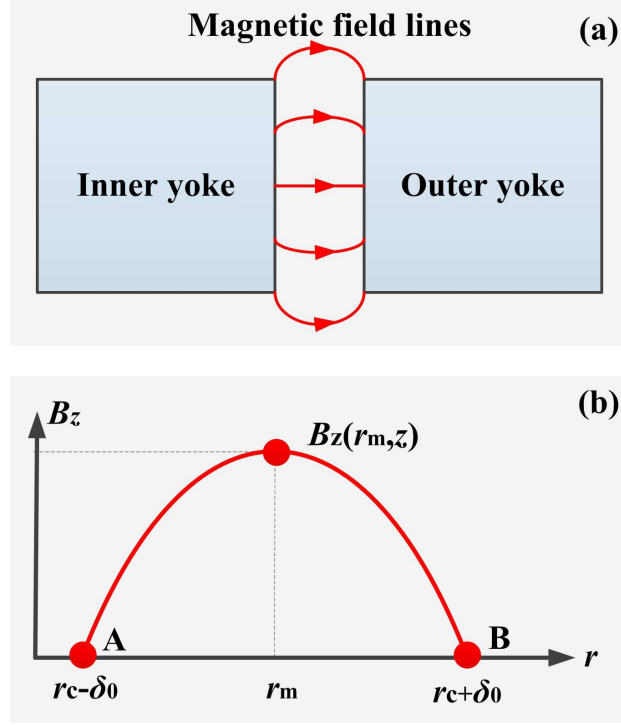}
\caption{(a) Schematic of the fringe field effect for the magnetic filed in the air gap. (b) A typical $B_z(r)$ profile along the radical direction between two surfaces of the inner and outer yokes at a certainty vertical position $z(z\neq 0)$.}
\label{Fig.2}
\end{figure}

In this section, some discussions on the fringe field effect of the air gap are prepared for the algorithm. Based on the electromagnetic knowledge, the vertical component of the magnetic flux density in the air gap, i.e., $B_z(r, z)$, is generated by a finite height of the air gap, which is known as the fringe filed effect. The schematic of the fringe field for the magnetic filed in the air gap has been shown in figure \ref{Fig.2}(a), and it is known that the fringe effect would bend the magnetic flux at both upper and lower ends of the air gap. As a result, the absolute value for the vertical magnetic component $B_z(r,z)$ will increase along a vertical axis departing the center position ($z=0$). In theory, if the upper and lower magnets are symmetrical, only the magnetic field line at the radical center has zero vertical magnetic component, i.e., $B_z(r,0)=0$.

As high-permeability yokes are used, the boundary on the yoke-air surface is very sharp, and it is reasonable to consider the magnetic flux at both yoke-air surfaces to be ideally radical, i.e., $B_z(r_c-\delta_0/2, z)=B_z(r_c+\delta_0/2, z)=0$ where $\delta_0$ is the air gap width. It can be proved in mathematics, as well as in conjunction with Maxwell's equations, the $B_z(r)$ function is a typical one-extremum function. A typical $B_z(r)$ function curve with $z\neq0$ has been shown in figure \ref{Fig.2}(b), which usually has two approximate zero values at surfaces and one extremum $B_z(r_m, z)$ around the central air gap.

\subsection{Algorithm}
The proposed algorithm employs $N+1$ measurements of $B_r(z)$ profiles at different radii, i.e., $B_r(r_i, z)$ where $i=1, 2, ..., N, N+1$. The central profile $B_r(r_c, z)$ ($r_c$ is the central radius of the air gap), conventionally being measured in the velocity mode of a watt balance, is included and indexed as $i=N+1$. The other $N$ measurements of $B_r(z)$ profiles indexed form 1 to $N$ are specially designed for supplying additional information for the algorithm. It should be noted that a larger number of $N$ will obtain a better estimation accuracy but it would obviously increase the complexity of the measurements, e.g., a multi-coil with $N$ upper coil and $N$ lower coils is required if the GC method is applied. The goal of this paper is to represent the whole air gap magnetic field by very small numbers of additional $B_r(z)$ measurements, e.g., $N=1$, with acceptable accuracy for relaxing the alignment.

In the calculation, $N+1$ measurements of $B_r(z)$ profiles are taken, and hence $N+1$ values of $\partial B_z(r,z)/\partial r$ along the radical direction of the air gap can be calculated by equation (\ref{eq.MaxII2D}). As discussed in section \ref{2.3}, it is reasonable to make the vertical component of the magnetic flux density at both yoke-air surfaces equal zero, i.e., $B_z(r_c-\delta_0/2, z)=B_z(r_c+\delta_0/2, z)=0$. When $N$ is a small number, the information with knowing $N+1$ differential values and 2 initial values at yoke-air surfaces for the $B_z(r)$ function, however, is still not enough for solving the magnetic flux density along the radical direction.

In order to make up the information lacking and ensure the field representation accuracy, a pre-estimation of $B_z(r)$ is applied. It has been pointed out in section \ref{2.3} and shown in figure \ref{Fig.2}(b) that the $B_z(r)$ function is a one-extremum function with two zero values at yoke-air surfaces. A polynomial estimation of $B_z(r)$ function should be feasible and can be applied in the algorithm.
The polynomial estimation employs the $B_r(r_c,z)$ profile and the other $N$ additional $B_r(z)$ measurements at different radii for best estimating the vertical magnetic component in the radical direction, which can yield a polynomial estimator with maximum order $K=N+2$ . The main idea of the proposed approach is to fit the $B_z(r)$ function at every vertical coordinate by a $N+2$ order polynomial function in dimension $r$ of the air gap as
\begin{equation}
B_z(r,z)=\alpha(z)(r-r_c+\frac{\delta_0}{2})(r-r_c-\frac{\delta_0}{2})\prod_{i=1}^{N}(r-\beta_i),
\label{eq.times}
\end{equation}
where $\alpha(z)$ is a gain factor; $\beta_i$ is the $(i+2)$th solution of $B_z(r,z)=0$ while $\beta_{-1}=r_c+\delta_0/2$ and $\beta_0=r_c-\delta_0/2$ are considered as its first and second solutions.

It can be seen from equation (\ref{eq.times}) that $B_z(r, z)$ is expressed by a product of two functions with separated directions, i.e., the two dimensions ($r$ and $z$) of the $B_z(r,z)$ function is decoupled, therefore, the calculation of the magnetic field in the air gap is greatly simplified. For each $B_r(r_i, z)$ measurement, it can be written by equation (\ref{eq.MaxII2D}) as
\begin{equation}
\frac{\partial B_r(r_i,z)}{\partial z}=\frac{\partial B_z(r_i,z)}{\partial r},~~~ i=1,2,...,N, c.
\label{eq.brm}
\end{equation}

The left part of equation (\ref{eq.brm}) is directly measured while the right part of equation (\ref{eq.brm}) can be expressed by the estimator following equation (\ref{eq.times}). As a result, $N+1$ unknown parameters, i.e., $\beta_i(i=1,2,...,N)$ and $\alpha$, can be solved by $N+1$ equations in (\ref{eq.brm}). Substituting the solved parameters back into equation (\ref{eq.times}), the $B_z(r,z)$ function on the whole air gap is obtained.

Knowing the $B_z(r,z)$ function in the air gap, we can directly calculate values of the differential function $\partial B_z(r,z)/ \partial z$ at any single point $(r,z)$. According to equation (\ref{eq.maxI2D}), $\partial [rB_r(r,z)]/\partial r$ is solved as
\begin{equation}
\frac{\partial [rB_r(r,z)]}{\partial r}=-r\frac{\partial B_z(r,z)}{\partial z}.
\label{eq.brd}
\end{equation}

To ensure the calculation accuracy in the actual operation area of a watt balance, the initial condition of $r_cB_r(r_c,z)$ is applied in the algorithm, the $rB_r(r,z)$ function and hence the $B_r(r,z)$ function in the air gap is solved.

In summary, both $B_z(r,z)$ and $B_r(r, z)$, i.e., a 3D magnetic field representation in the air gap, can be solved by polynomial estimation of $B_z(r)$ function. The minimum number of additional $B_r(z)$ measurements for the estimation equation (\ref{eq.times}) in theory can be zero, i.e., $N=0$. Note that this case is established only when a $B_r(z)$ profile with $r\neq r_c$ is known. For watt balance experiments, the $B_r(r_c, z)$ profile is preferred to obtain a larger vertical interval with good uniformity but the gain factor $\alpha$ in this case cannot be solved. Therefore, in order to represent the 3D magnetic field in the air gap for watt balance, at least one additional $B_r(z)$ profile measurement should be taken.

\section{Numeral verification}
\subsection{Simulation setup}
\begin{figure}
\center
\includegraphics[width=2.8in]{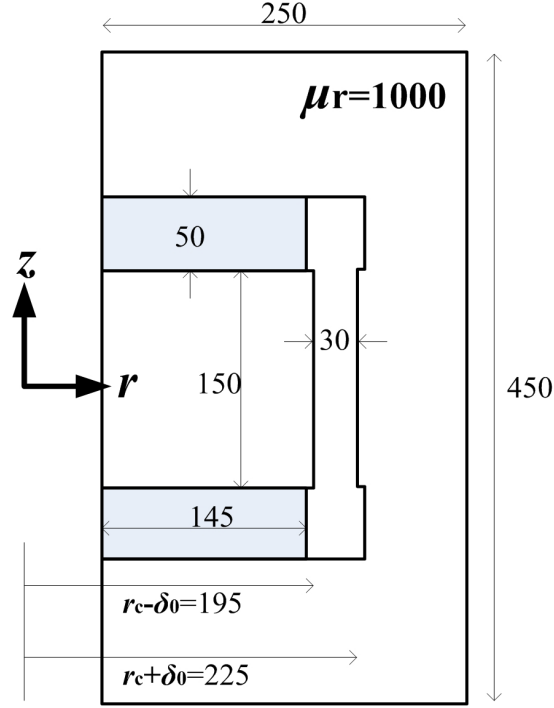}
\caption{Parameters setup of the numeral simulation (unit:mm).}
\label{Fig.4}
\end{figure}

In order to evaluating the model accuracy presented in section \ref{section2}, here some numeral simulations based on finite element method (FEM) are performed. In the simulation, the geometrical parameters of the magnet have been shown in figure \ref{Fig.4}. The relative permeability of the yoke is set as $\mu_r=1000$ and the magnetic strength of the permanent magnet is set as 800kA/m in the vertical $z$ direction. Note that in the simulation of a watt balance magnet, the nonlinearity of the yoke permeability, instead of a constant number, should be considered. However, in our approach, the nonlinear information has already been contained in measurements of $B_r(r_i, z)$ profiles. In such cases, the real magnet can be simplified by a constant-permeability one and no obvious error would be generated in evaluating the model accuracy.
In the simulation, five magnetic profiles $B_r(219$mm$,z)$, $B_r(215$mm$,z)$, $B_r(210$mm$,z)$, $B_r(205$mm$,z)$, and $B_r(201$mm$,z)$ shown in figure \ref{Fig.5}, are firstly calculated as possible known conditions of the presented model to simulated the actual measurements of either a GC coil or a magnetic probe.

\subsection{A case with $N=1$}
\begin{figure}
\center
\includegraphics[width=3.8in]{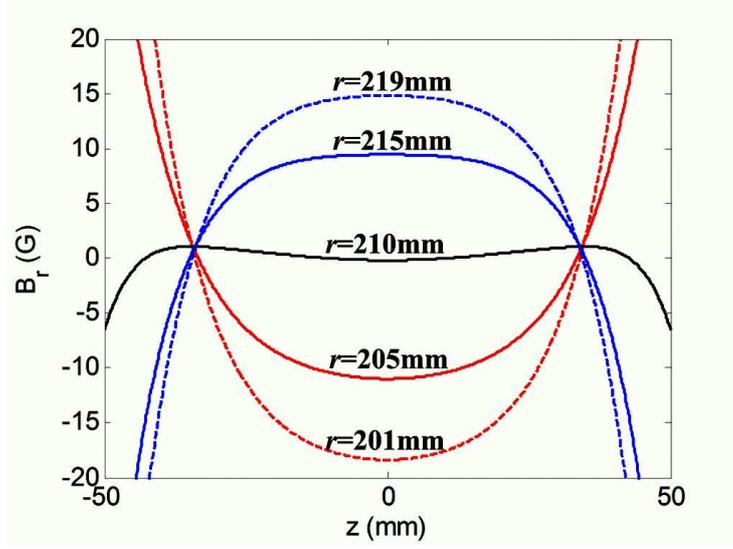}
\caption{$B_r(z)$ profiles applied in the calculation of the presented model.}
\label{Fig.5}
\end{figure}

Here the simplest condition with $N=1$ additional $B_r(z)$ profile when the third-order polynomial estimator is calculated as an example. In this case, $B_z(r,z)$ is considered to be a cubic function of $r$ with arbitrary $z$ position. Following equation (\ref{eq.times}), $B_z(r,z)$ function can be expressed as
\begin{equation}
B_z(r,z)=\alpha(z)(r-r_c+\frac{\delta_0}{2})(r-r_c-\frac{\delta_0}{2})(r-\beta_1),
\end{equation}
 and its partial derivative is written as
 \begin{equation}
 \frac{\partial B_z(r,z)}{\partial r}=2(r-r_c)(r-\beta_1)\alpha(z)+[(r-r_c)^2-\frac{\delta_0^2}{4}]\alpha(z).
 \label{eq.dadd}
 \end{equation}
 Using the profile $B_r(r_c, z)$, $\alpha(z)$ can be solved as
 \begin{equation}
 \alpha(z)= -\frac{4}{\delta_0^2}\frac{\partial B_r(r_c,z)}{\partial z}.
 \label{eq.A}
 \end{equation}

\begin{figure}
\center
\includegraphics[width=5.8in]{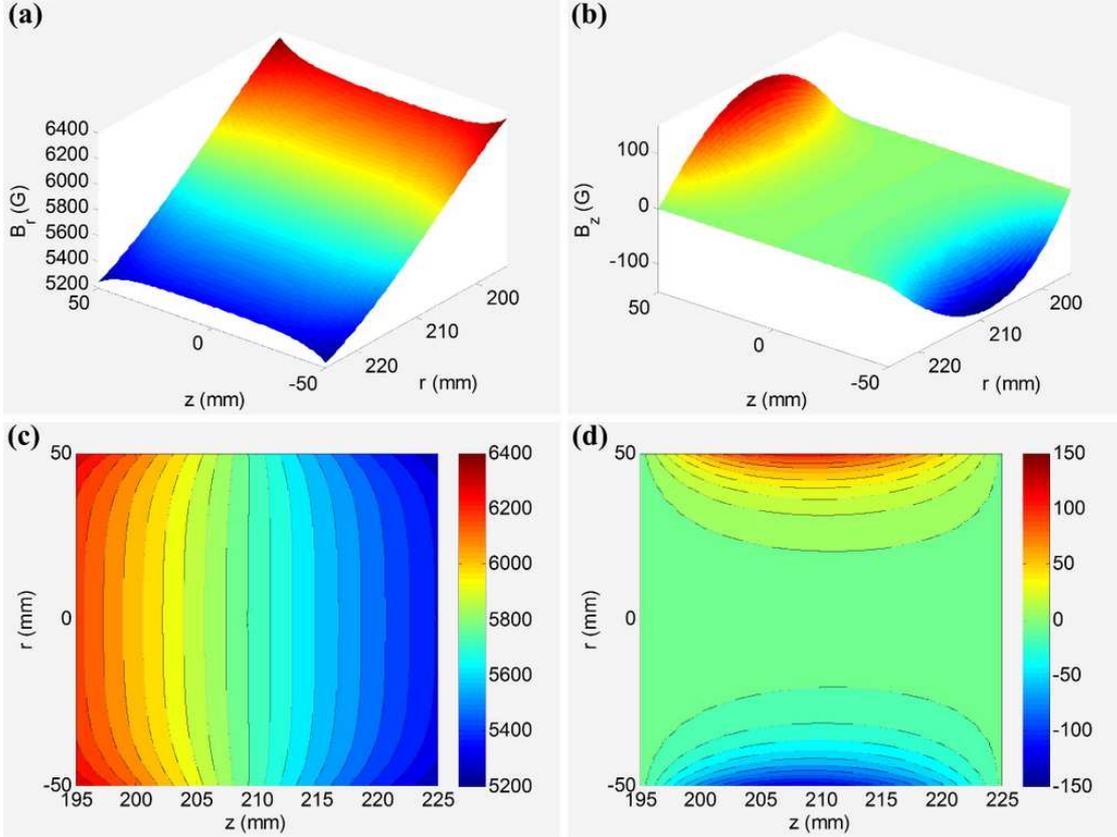}
\caption{Calculation results of magnetic flux density distribution with vertical profiles $B_r(210$mm$, z)$ and $B_r(201$mm$, z)$. (a) Magnetic flux density distribution of $B_r(r,z)$ in 3D. (b) Magnetic flux density distribution of $B_z(r,z)$ in 3D. (c) Top view of the magnetic flux density distribution $B_r(r,z)$. (d) Top view of the magnetic flux density distribution $B_z(r,z)$.}
\label{Fig.6}
\end{figure}

Here an additional measurement $B_r(r_c+\delta,z)$ where $\delta\in(-\delta_0/2,\delta_0/2)$ is taken. Based on equation (\ref{eq.dadd}), it can be obtained that
 \begin{equation}
 \frac{\partial B_r(r_c+\delta,z)}{\partial z}=\delta(2r_c+3\delta-2\beta_1)\alpha(z)-\frac{\delta_0^2}{4}\alpha(z).
 \label{eq.B}
 \end{equation}

Substituting equation (\ref{eq.A}) into equation (\ref{eq.B}), $\beta_1$ can be calculated as
  \begin{equation}
\beta_1=r_c+\frac{3\delta}{2}-\frac{\delta_0^2}{8\delta}[1-\frac{\partial B_r(r_c+\delta,z)/\partial z}{\partial B_r(r_c,z)/\partial z}].
 \end{equation}

In the calculation, only two $B_r(z)$ profiles, $B_r(r_c,z)$ and $B_r(r_1,z)$ shown in figure \ref{Fig.5}, are applied. An area with $r\in(195$mm$,225$mm$)$, $z\in(-50$mm$,50$mm$)$, where the watt balance is operated, has been focused. The calculation results of $B_r(r,z)$ and $B_z(r,z)$ with $r_c=210$mm and $r_1=201$mm have been shown in figure \ref{Fig.6}.

\begin{figure}
\center
\includegraphics[width=5.6in]{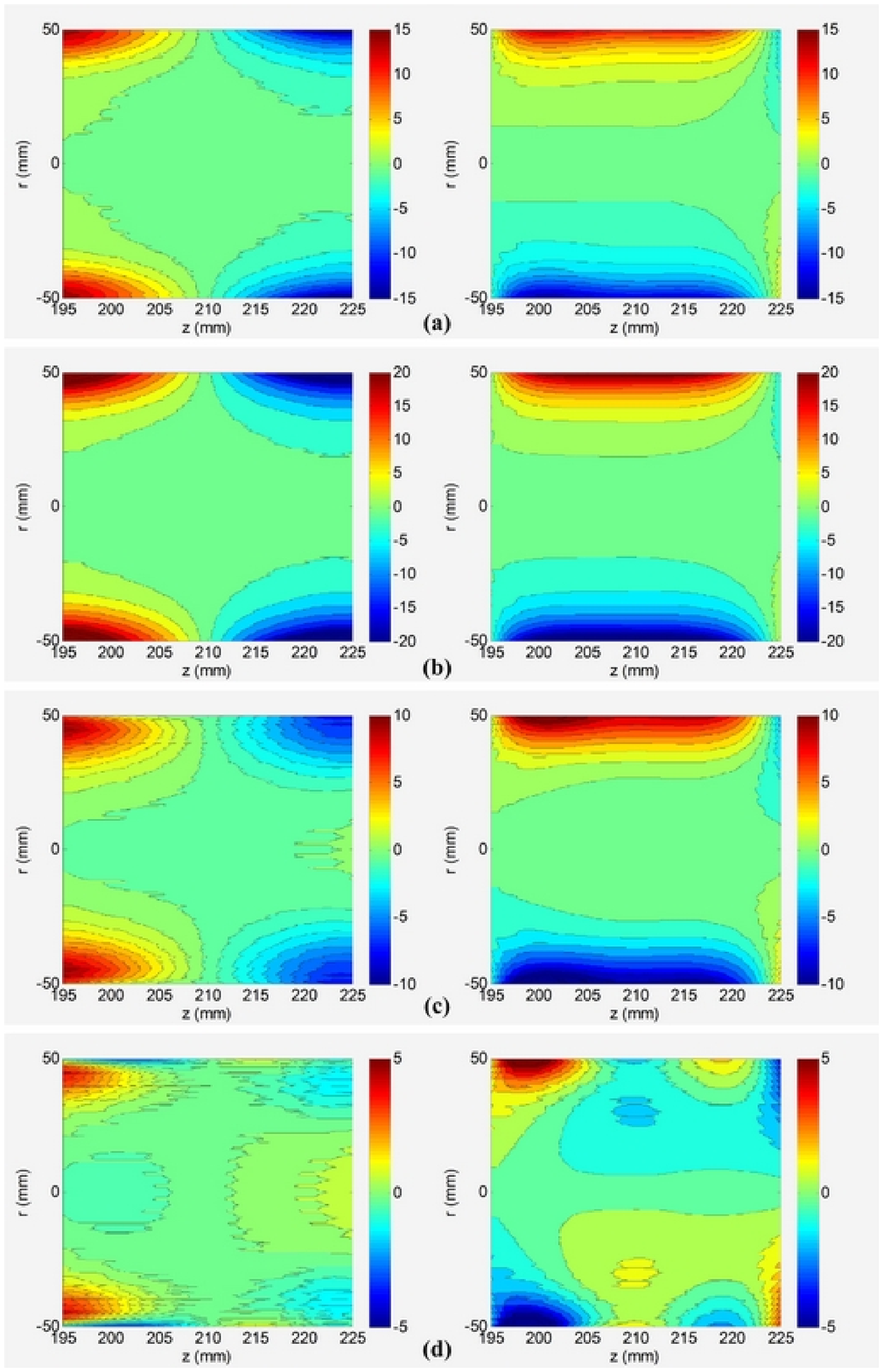}
\caption{Filed calculation error in Gauss when $N=1$. The additional $B_r(z)$ profile is adopted at the radius $r_1$ that (a) $r_1=201$mm, (b) $r_1=205$mm, (c) $r_1=215$mm, and (d) $r_1=219$mm. In each subgraph, the left is the magnetic flux density error of $B_r(r,z)$ while the right is the magnetic flux density error of $B_z(r,z)$.}
\label{Fig.7}
\end{figure}

It can be seen from the calculation results that the amplitude of $B_r(r)$ decays along $r$ direction approximately following a $1/r$ relation. This relationship can be physically explained: the total flux through the air gap, $\Phi$, is a fixed number only determined by the MMF of the permanent magnet and the magnetic reluctance of the circuit; since the magnetic flux on each surface is approximate uniform, the total flux can be expressed as $\Phi=2\pi rdB(r)$ where $d$ is the height of the air gap. Therefore, $B_r(r)=\Phi/(2\pi rd)$, i.e., $B_r(r)$ decays along $r$ direction with $1/r$. Besides, the calculation of $B_z(z)$ also clearly shows the fringe filed effect, i.e., $B_z$ increases quickly in the air gap when $z$ is departing from the central radical surface $z=0$.

In order to show the calculation accuracy of the proposed algorithm, the errors in the focused space, compared to FEM simulations, are calculated.
The error maps for both $B_r(r,z)$ and $B_z(r,z)$ with different choices of the additional $B_r(r_1,z)$ profile, i.e., $r_1$ respectively equals to 201mm, 205mm, 215mm, and 219mm, have been shown in figure \ref{Fig.7}.
It can be seen from figure \ref{Fig.7} that the calculation error is closely related to the choice of the additional $B_r(r_1, z)$ profile.
The maximum relative error in different cases when $N=1$ is about 0.1\%--0.4\% for $B_r(r,z)$ and about 4\%--16\% for $B_z(r,z)$.
The calculation result shows that the error when the additional $B_r(z)$ is taken at $r_1>r_c$ is smaller than that when the additional $B_r(z)$ is taken at $r_1<r_c$.
Therefore, it is suggested to take an additional $B_r(z)$ measurement in the outer half of the air yoke, e.g., $r=r_c+2\delta_0/3$, when a cubic estimator is applied.

\subsection{High order estimators}
\begin{figure}
\center
\includegraphics[width=5.6in]{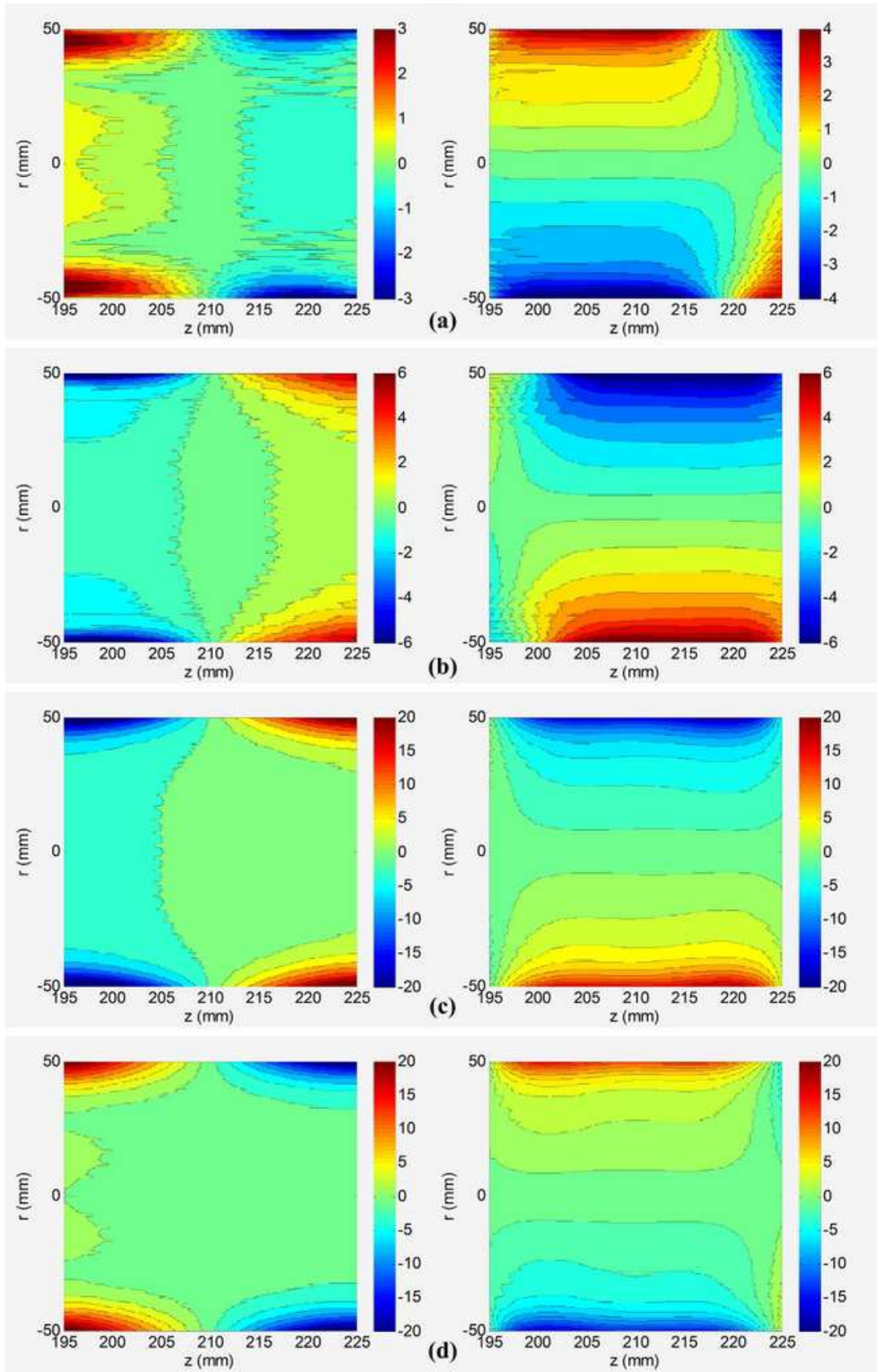}
\caption{Filed calculation error in Gauss when $N=2$. Additional $B_r(z)$ profiles are adopted at two radii, $r_1$ and $r_2$. (a) $r_1=201$mm, $r_2=205$mm; (b) $r_1=215$mm, $r_1=219$mm; (c) $r_1=205$mm, $r_2=219$mm; (d) $r_1=215$mm, $r_2=201$mm. In each subgraph, the left is the magnetic flux density error of $B_r(r,z)$ while the right is the magnetic flux density error of $B_z(r,z)$.}
\label{Fig.8}
\end{figure}

Another simulation example when $N=2$ is also taken and the simulation results have been shown in figure \ref{Fig.8}. Two valuable conclusions have been found from the calculation: Firstly, two additional measurements, $B_r(r_1, z)$ and $B_r(r_2, z)$, can not be set symmetrical on $r_c$, i.e., $r_1-r_c\neq r_c-r_2$. Otherwise, an infinite value for the gain factor $\alpha$ would obtained. In this case, $\alpha$ is expressed as
\begin{equation}
\alpha(z)=\frac{\delta_0^2\delta_1\frac{\partial [B_r(r_c,z)-B_r(r_2,z)]}{\partial z}-\delta_0^2\delta_2\frac{\partial [B_r(r_c,z)-B_r(r_1,z)]}{\partial z}+12\delta_1\delta_2(\delta_1-\delta_2)\frac{\partial B_r(r_c,z)}{\partial z}}{4\delta_0^2\delta_1\delta_2(\delta_1-\delta_2)(\delta_1+\delta_2)},
\end{equation}
where $\delta_1=r_1-r_c$ and $\delta_2=r_2-r_c$. Obviously, the condition $\delta_1+\delta_2=0$ is not allowed.
Secondly, the calculation error with $r_1$ and $r_2$ chosen in the same half side of the air gap, either outer or inner parts, is smaller than the cases with $r_1$ and $r_2$ in different halves of the air gap.
Therefore, if the measurement with $N=2$ is taken, it is better to choose a same half of the air gap in radical direction, e.g., $r_1=r_c-\delta_0/3$ and $r_2=r_c-2\delta_0/3$, or $r_1=r_c+\delta_0/3$ and $r_2=r_c+2\delta_0/3$.

Higher order estimators with $N(N>2)$ measurements of $B_r(z)$ profiles contain more filed information, and in theory can improve the calculation accuracy. However, it would synchronously increase the complexity for both measurement and analytical solving unknown parameters of the estimator. Limited to the paper length, no further calculation with higher order estimators is presented.

\section{Discussion}
The misalignment error $\varsigma$ of a watt balance has been presented in \cite{Robinson2012alignment}, expressed as
\begin{equation}
\varsigma=\frac{UI}{F_zv_z}-1=\frac{F_xv_x}{F_zv_z}+\frac{F_yv_y}{F_zv_z}+\frac{\Gamma_x\omega_x}{F_zv_z}
+\frac{\Gamma_y\omega_y}{F_zv_z}+\frac{\Gamma_z\omega_z}{F_zv_z},
\label{eq.error}
\end{equation}
where $F=(F_x, F_y, F_z)$ and $\Gamma=(\Gamma_x, \Gamma_y, \Gamma_z)$ are magnetic forces and torques generated in the weighing mode. Note that $F$ and $\Gamma$ are results of the interaction between the coil current $I$ and the magnetic flux $\psi$, which can be written respectively as
\begin{equation}
F=-I({\bf i}\frac{\partial \psi}{\partial x}+{\bf j}\frac{\partial \psi}{\partial y}+{\bf k}\frac{\partial \psi}{\partial z}),
\end{equation}
\begin{equation}
\Gamma=-I({\bf i}\frac{\partial \psi}{\partial \theta_x}+{\bf j}\frac{\partial \psi}{\partial \theta_y}+{\bf k}\frac{\partial \psi}{\partial \theta_z}).
\end{equation}
In above equations, $\theta=(\theta_x, \theta_y, \theta_z)$ is the rotation angle of coil, and $({\bf i}, {\bf j}, {\bf k})$ is a unit vector in $(x, y, z)$ space.

Base on measurements of $B_r(z)$ profiles and the presented model, 3D functions of the magnetic flux in the magnet air gap $\psi=(\psi_x, \psi_y, \psi_z)$ can be calculated, and hence the magnetic forces $F=(F_x, F_y, F_z)$ and torques $\Gamma=(\Gamma_x, \Gamma_y, \Gamma_z)$ as functions of geometrical space are known. Further, in conjunction with optical measurements of the velocity $v=(v_x, v_y, v_z)$ and the angular velocity $\omega=(\omega_x, \omega_y, \omega_z)$, the misalignment error $\varsigma$ in equation (\ref{eq.error}) can be determined, and a correction of misalignment errors in theory can be applied.

It should be noted that all terms on the right side of equation (\ref{eq.error}) are tinny values compared to 1, which conventionally should be adjusted below $1\times10^{-8}$ by alignment. As presented, the proposed algorithm is expected to represent the 3D magnetic profile with an accuracy of several parts in $10^2$ with few additional measurements of $B_r(z)$ profiles. On the one hand, under such situations, a whole field parameter can be supplied for alignment analysis. On the other hand, if the field information is used for misalignment correction, the alignment would be relaxed, e.g., several parts in $10^{7}$.

\section{Conclusion}
An algorithm employing a polynomial estimator for functional fit of $B_z(r)$  is presented to characterize the 3D magnetic field of a watt balance. Based on Maxwell's equations, the relations of two magnetic components in air gap of a watt balance magnet, $B_r(r,z)$ and $B_z(r,z)$, are directly modelled by decoupling the $B_z(r,z)$ function in $r$ and $z$ dimensions. The accuracy of the presented model is evaluated by comparisons to the FEM simulation, and the calculation results showed the alignment of a watt balance experiment can be greatly relaxed by few additional $B_r(z)$ profiles. As the model relies on real measurements of $B_r(z)$ functions, the nonlinear information of the yokes has been contained in the calculation, and in theory can be applied in watt balances with different magnetic profiles. The model supplies basic magnetic field parameters, which can be used for alignment and misalignment corrections.

\section*{Acknowledgement}
The idea is inspired by the GC measurement of the magnetic profile for NIST-4 magnet. The authors would like to thank Prof. Zhonghua Zhang for valuable discussions and Mr. Zhuang Fu for helping us to find analytical solutiona of the $N=2$ case by Mathematica.

\end{document}